\documentclass[nofootinbib,superscriptaddress,twocolumn,showpacs,preprintnumbers,amsmath,amssymb]{revtex4-1}

\usepackage{graphicx}
\usepackage{grffile}
\usepackage{slashed}
\usepackage{bbm}
\usepackage{footmisc}
\usepackage{multirow}

\usepackage{hyperref}
\usepackage{amssymb}
\usepackage{amsmath}
\usepackage{color}
\usepackage{xspace}
\usepackage{caption}
\usepackage{subcaption}
\usepackage{natbib}

\usepackage{datetime}
\usepackage{color,fancybox}

\newcommand{\MeV}{\ensuremath{\,\mathrm{MeV}}\xspace}
\newcommand{\GeV}{\ensuremath{\,\mathrm{GeV}}\xspace}
\newcommand{\TeV}{\ensuremath{\,\mathrm{TeV}}\xspace}

\newcommand{\fb}{\ensuremath{\,\mathrm{fb}}\xspace}

\newcommand{\order}[1]{\mathcal{O}\!\left(#1\right)}

\newcommand{\mat}[1]{\ensuremath{\begin{pmatrix}#1\end{pmatrix}}}
\newcommand{\bea}{\begin{eqnarray}}
\newcommand{\eea}{\end{eqnarray}}

\newcommand{\eq}[1]{Eq.~(\ref{#1})}
\newcommand{\bib}[1]{Ref.~\cite{#1}}
\newcommand{\fig}[1]{Fig.~\ref{#1}}
\newcommand{\tab}[1]{Table~\ref{#1}}
\newcommand{\sect}[1]{Section~\ref{#1}}

\begin{document}
\title{Next-to-leading order QCD corrections to $W^+W^+$ and $W^-W^-$ production in association with two jets}

\preprint{FTUV-13-2211\;\; IFC/13-89\;\; KA-TP-39-2013\;\;LPN13-096\;\;SFB/CPP-13-106}

\author{Francisco~Campanario}
\email{francisco.campanario@ific.uv.es}
\affiliation{Theory Division, IFIC, University of Valencia-CSIC, E-46980
  Paterna, Valencia, Spain}
\author{Matthias~Kerner}
\email{matthias.kerner@kit.edu}
\affiliation{Institute for Theoretical Physics, KIT, 76128 Karlsruhe, Germany}
\author{Le~Duc~Ninh}
\email{duc.le@kit.edu}
\affiliation{Institute for Theoretical Physics, KIT, 76128 Karlsruhe, Germany}
\affiliation{Institute of Physics, Vietnam Academy of Science and Technology, \\
10 Dao Tan, Ba Dinh, Hanoi, Vietnam}
\author{Dieter~Zeppenfeld}
\email{dieter.zeppenfeld@kit.edu}
\affiliation{Institute for Theoretical Physics, KIT, 76128 Karlsruhe, Germany}

\begin{abstract}
We present a study of $W^+W^+jj$ and $W^-W^-jj$ production including leptonic decays in hadron-hadron collisions. 
The full electroweak and QCD induced contributions and their interferences are calculated at leading order. 
We find that, for inclusive cuts, the interference effects can be large if the jets are produced with large transverse momentum where, however, the production rate is suppressed. 
We also discuss the vector-boson-fusion (VBF) cuts and show the validity of the VBF approximation. 
The NLO QCD corrections to the QCD-induced channels are also calculated. 
Compared to the previous calculation, we allow the intermediate $W$ bosons to be off-shell. For on-shell $W$ production, we obtain an excellent agreement with previous results.  
Our code will be publicly available as part of 
the parton level Monte Carlo program {\texttt{VBFNLO}}. 
\end{abstract}


\maketitle

\section{Introduction}
The production processes of two vector bosons in association with two jets at the LHC are important 
since they allow to probe vector boson scattering and are sensitive to triple and quartic gauge couplings. 
In addition, they are also backgrounds to various searches for beyond the standard model physics. 
Their production can be classified into two groups, namely the electroweak (EW) induced channels 
of order $\order{\alpha^6}$ and the QCD-induced processes of order $\order{\alpha_s^2 \alpha^4}$. 
Moreover, the EW contributions are divided into the $t$- and $u$-channels named 
``vector boson fusion'' (VBF) mechanisms and the $s$-channel corresponding to 
the production of three EW gauge bosons with one off-shell gauge boson decaying into a quark-antiquark pair. 
The VBF processes, including in particular $VV\rightarrow VV$ scattering, have been calculated at NLO QCD in Refs.~\cite{Jager:2006zc,Jager:2006cp,Bozzi:2007ur,Jager:2009xx,Denner:2012dz} for all combinations of massive gauge bosons. 
A similar calculation with a $W$ boson and a real photon in the final state
has been done in \bib{Campanario:2013eta}. NLO QCD corrections to triboson
production including leptonic decays were computed in 
Refs.~\cite{Hankele:2007sb,Campanario:2008yg,Bozzi:2009ig,Bozzi:2010sj,Bozzi:2011en,Bozzi:2011wwa}
and are available via the {\texttt{VBFNLO}}
program~\cite{Arnold:2008rz,*Arnold:2012xn} (see also 
Refs.~\cite{Lazopoulos:2007ix,Binoth:2008kt,Baur:2010zf} for on-shell
production and \bib{Nhung:2013jta} for NLO EW corrections).\\[2mm]

NLO QCD calculations of the QCD-induced processes involving nontrivial color structures and the calculation of up to six-point one-loop integrals are more challenging. 
These calculations have been done for $W^+W^+jj$~\cite{Melia:2010bm}, $W^+W^-jj$~\cite{Melia:2011dw,Greiner:2012im}, $W^\pm Zjj$~\cite{Campanario:2013qba} 
and $\gamma \gamma jj$~\cite{Gehrmann:2013bga} production. The calculations of Refs.~\cite{Jager:2009xx,Melia:2010bm} have been implemented in the POWHEG BOX framework~\cite{Alioli:2010xd} as described in Refs.~\cite{Melia:2011gk,Jager:2011ms} for the $W^+W^+jj$ channel. 
This implementation showed that the computing-time cost due to the calculation of the virtual 
corrections to the QCD-induced processes is a practical bottleneck~\cite{Melia:2011gk}. \\[2mm]

The signature with same-charge $W$ bosons is interesting because the backgrounds are small and 
it is related to the issue of double-parton scatterings. 
In this paper, we provide another independent calculation of the NLO QCD corrections to the 
QCD-induced $W^+W^+jj$ and $W^-W^-jj$ processes. 
Compared to the previous study presented in Ref.~\cite{Melia:2010bm}, we also 
include the off-shell gauge boson contribution and the gluon self-energy correction with a top quark in the loop. 
The importance of interference effects between the EW and QCD induced channels is a frequently asked question 
when studying vector boson pair production in association with two jets. 
These interferences are most important for the processes considered here due to 
the absence of gluon induced processes at leading order (LO) and since only 
left-chiral quarks contribute to the EW as well as QCD mechanisms. 
We discuss these effects at LO and expect that the results can be used as an upper limit for other $VVjj$ processes.
We also aim at having a very fast code to solve the above computing-time problem. 
The code will be publicly available as part of the parton level Monte Carlo program {\texttt{VBFNLO}}. \\[2mm]

This paper is organized as follows. In the next section, details of our calculation and code implementation 
are given. Numerical results are presented in \sect{sec:results} and the conclusions in the last section. Finally, 
in the appendix we provide results at the amplitude squared level 
at a random phase-space point to facilitate comparisons with our results.

\section{Computational details}
\label{sec:comp_details}
As always done in the {\texttt{VBFNLO}} program, 
the leptonic decays of the EW gauge bosons are consistently included, with all off-shell effects 
and spin correlations taken into account. In the following, 
we consider the specific leptonic final state $e^+\nu_e\mu^+\nu_\mu$ 
and $e^-\bar{\nu}_e\mu^-\bar{\nu}_\mu$. The total results for all possible decay
channels ({\it i.e.} $e^+\nu_e\mu^+\nu_\mu$, $e^+\nu_ee^+\nu_e$, $\mu^+\nu_\mu\mu^+\nu_\mu$ 
in the $W^+W^+jj$ case and accordingly for the $W^-W^-jj$ production) 
can, apart from negligible identical lepton interference effects, 
be obtained by multiplying our predictions by a factor two. 
For simplicity, we choose to describe the resonating $W^\pm$ propagators 
with a fixed width and keep the weak-mixing angle real.

\begin{figure}[ht]
  \centering
  \includegraphics[width=0.80\columnwidth]{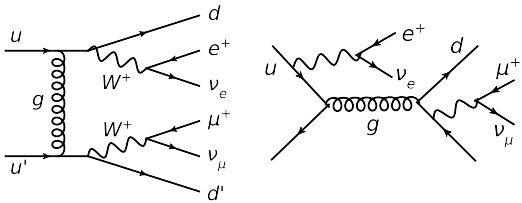}
  \caption{Representative tree-level Feynman diagrams of the QCD-induced mechanisms 
for the process $pp\rightarrow e^+\nu_e \mu^+ \nu_\mu~jj$.}
  \label{fig:Born}
\end{figure}

\begin{figure}[ht]
   \centering
  \includegraphics[width=0.9\columnwidth]{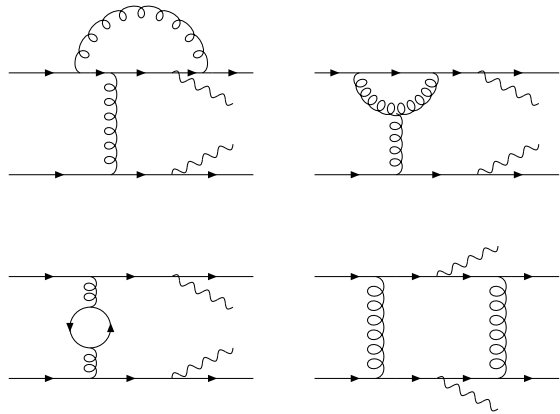}\caption{Diagram types contributing to the virtual amplitude.}
 \label{fig:classes}
\end{figure}

A special feature of this process is that there are no subprocesses with external 
gluons at LO. All tree-level Feynman diagrams have two quark lines with a $W$ boson 
attached to each, as displayed in \fig{fig:Born}. As a consequence, the total 
cross section without any cuts is finite at LO. 
Crossing symmetry is used to obtain, for two generations of quarks, all $20$ subprocesses from the minimal set 
of two generic subprocesses (one subprocess involves same-generation 
quarks such as $u\bar{d}\to \bar{u}d W^+W^+$ and the other with different-generation 
quarks such as $u\bar{d}\to \bar{c}s W^+W^+$). It is 
obvious that the subprocesses with same-generation quarks are more complicated and 
include all diagrams of the different-generation subprocesses as a half set. This feature is 
used to avoid calculating the same diagrams twice. At NLO, there are the virtual and the real 
corrections. \fig{fig:classes} shows some selected contributions to the virtual amplitude, which involves,
in particular, the hexagon diagrams. The most difficult part of the calculation is computing the virtual amplitudes with up to six-point rank-four one-loop 
tensor integrals. 
There are eight six-point diagrams for each of four independent subprocesses with 
same generation quarks. For different generation quarks there are four six-point diagrams for six independent subprocesses.
The calculation of tensor 
integrals is done using Passarino-Veltman reduction~\cite{Passarino:1978jh} 
for up to $4$-point diagrams and the method of \bib{Denner:2005nn} (see also Refs.~\cite{Campanario:2011cs,Binoth:2005ff}) for higher-point tensor integrals. 
The scalar integrals are calculated as in Refs.~\cite{'tHooft:1978xw, Bern:1993kr, Dittmaier:2003bc, Nhung:2009pm, Denner:2010tr}. 
The real emission contribution includes, for two generations of quarks, $36$ subprocesses with seven particles in the final state. 
After removing the UV divergences in the virtual amplitude by the renormalization of $\alpha_s$, both the virtual and real corrections are separately infrafred divergent. These divergences cancel in the sum for 
infrared-safe observables such as the inclusive cross section and jet distributions.  
We use the dimensional regularization method~\cite{'tHooft:1972fi} 
to regularize the UV and the infrared divergences and use an anticommuting prescription of $\gamma_5$~\cite{Chanowitz:1979zu}. 
The virtual and real emission contributions are combined using the 
Catani-Seymour dipole subtraction algorithm~\cite{Catani:1996vz}.   

We have constructed two independent implementations of the above described method. 
The results of the two computer codes are in full agreement, typically $10$ to $12$ digits 
with double precision, at the amplitude level for all subprocesses at NLO. The integrated part of the 
dipole subtraction term in~\bib{Catani:1996vz} has been compared at the integration level. 
Moreover, we have also compared to the results of Ref.~\cite{Melia:2010bm} using their 
settings and found a very good agreement. 
The first implementation is done in the {\texttt{VBFNLO}} framework~\cite{Arnold:2008rz,*Arnold:2012xn}, which will be described below. 
The second implementation uses {\texttt{FeynArts-3.4}}~\cite{Hahn:2000kx} 
and {\texttt{FormCalc-6.2}}~\cite{Hahn:1998yk} to obtain the virtual amplitudes. 
The scalar and tensor one-loop integrals are evaluated with the in-house 
library {\texttt{LoopInts}}. The tree-level amplitudes for both LO and NLO real emission contributions 
are calculated in an optimized way using {\texttt{HELAS}}~\cite{Murayama:1992gi,Alwall:2007st} routines.    

\begin{figure}[t!]
  \centering
  \includegraphics[width=0.9\columnwidth]{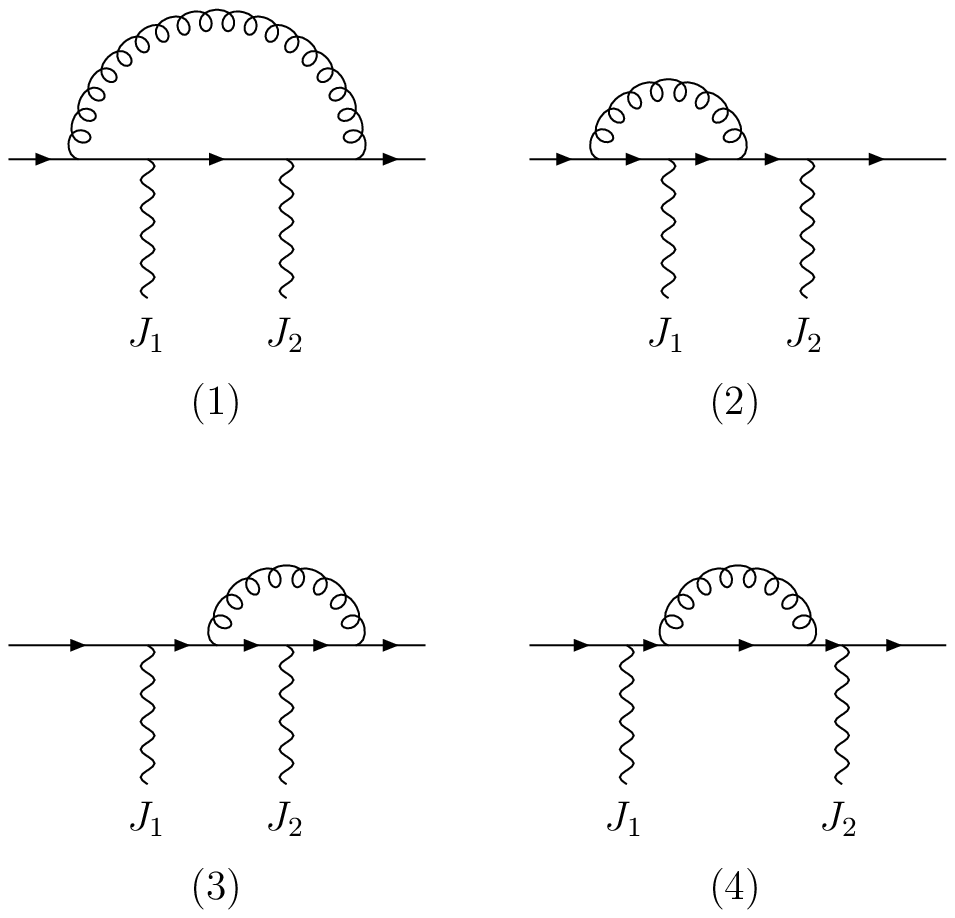}\caption{Contributions of the abelian boxline.}
 \label{fig:boxabe}
  \end{figure}

In the following, we sketch the main implementation which has been added to the 
{\texttt{VBFNLO}} program and will be made public. 
We use the spinor-helicity formalism of \bib{Hagiwara:1988pp} throughout the
code. The virtual amplitudes can be classified into four different topologies as depicted in
Fig.~\ref{fig:classes}. They are evaluated with the use of five generic
building blocks as described below. 
The upper diagrams of Fig.~\ref{fig:classes}, which contain loop
corrections to a quark line and are often called `bosonic' contributions,  
include sets of loop corrections to Born topologies with a fixed number and a fixed order of
external particles. They are classified into abelian and non-abelian
contributions. The first ones are computed with the so-called ``boxline''
depicted in \fig{fig:boxabe}. 
We use the effective current approach, thus, $J_1$ and $J_2$ should be understood as generic off-shell
currents which can be either a $W$ boson including the leptonic decays or a 
gluon connecting to the second quark line. The color factors associated with the
individual diagrams are proportional to $C_F-C_A/2$ or $C_F$ depending on whether the effective
gluon is attached to the loop. For example, if $J_1$ is a gluon then it is $C_F-C_A/2$ for 
the first and second diagrams and $C_F$ for the other ones. 

\begin{figure}[t]
   \centering
  \includegraphics[width=0.9\columnwidth]{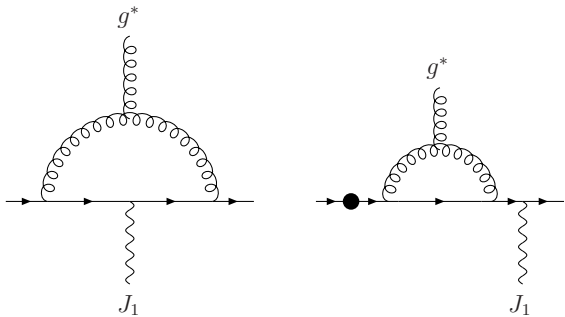}\caption{Nonabelian
    boxline contributions. The dot indicates the additional position, where the external boson can be attached.}
 \label{fig:boxnoabe}
\end{figure}
\begin{figure}[t]
   \centering
  \includegraphics[width=0.6\columnwidth]{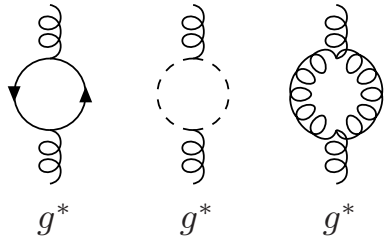}\caption{Self-energy
  contributions to the amplitude.}
 \label{fig:self}
\end{figure}

Corrections to a quark line containing non-abelian triple vertices are
computed with the so-called ``BoxlineNoAbe'' building block depicted 
in \fig{fig:boxnoabe}. $g^*$ and $J_1$ represent again generic effective
currents to which the building block can be contracted. These building
blocks have been extensively
checked in~\bib{Campanario:2011cs} and used in other {\texttt{VBFNLO}} processes.
The self-energy corrections, 
which are illustrated in the bottom left diagram of \fig{fig:classes} and include various 
contributions shown in \fig{fig:self}, form another building block. We note that 
the top-loop corrections are included.  

\begin{figure}[t]
   \centering
  \includegraphics[width=0.9\columnwidth]{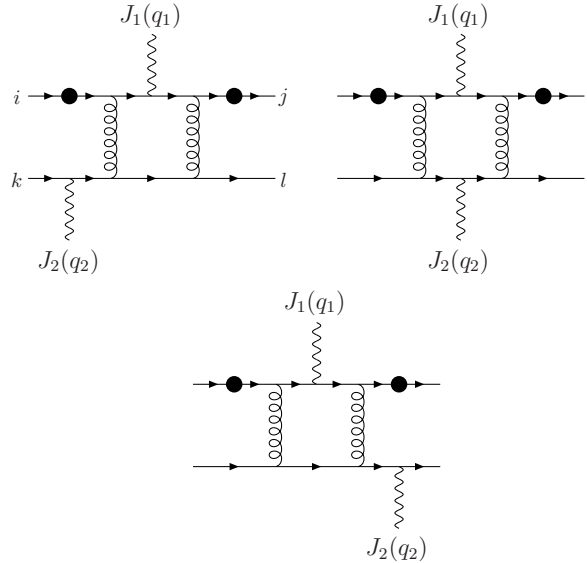}\caption{Hexbox contributions. The dots indicate additional positions, where the external boson on the upper line has to be attached.}
 \label{fig:hexbox2}
\end{figure}
Finally, there are building blocks involving from hexagon to box corrections with two quark lines directly 
attached to the loop as illustrated in the bottom right diagram of \fig{fig:classes} and further shown 
in \fig{fig:hexbox2}. 
They are gathered in EW gauge invariant subsets and named ``hexbox'' contributions. 
The first subset depicted in \fig{fig:hexbox2} consists of nine diagrams. 
The other subset is obtained by crossing the two gluon lines and constitutes an independent group.
To compute them, we generalized the software developed in~\bib{Campanario:2011cs} to be able to compute hexagon diagrams with two 
fermion lines. We use the Chisholm identities (see e.g. \bib{Sirlin:1981pi}), which reduces the CPU time required to evaluate 
the hexbox contributions by a factor ten.

While the color factors of the boxline and self-energy contributions are given by a constant times the color structure of the corresponding Born diagram, the hexbox diagrams involve a more complicated structure.
For the diagrams shown in Fig.~\ref{fig:hexbox2}, it is given by
\begin{equation}
\begin{aligned}
  (T^bT^a)_{ji}(T^bT^a)_{lk} &=-\frac{1}{6}T^a_{ji}T^a_{lk}+\frac{1}{2}T^a_{li}T^a_{jk}\\ &= -\frac{1}{6}\mathcal{C}_t+\frac{1}{2}\mathcal{C}_u,
\end{aligned}
\end{equation}
where $\mathcal{C}_t$ and $\mathcal{C}_u$ are the color structures of the corresponding $t$ and $u$ channel Born diagrams. The color factors of the hexbox diagrams with crossing gluons are
\begin{equation}
 (T^aT^b)_{ji}(T^bT^a)_{lk} = \frac{4}{3}\mathcal{C}_t+\frac{1}{2}\mathcal{C}_u.
\end{equation}
After adding each Feynman diagram to the corresponding color structure(s), the squared amplitude can be calculated using
\begin{equation}
\mat{\mathcal{C}_t \\ \mathcal{C}_u}^*\cdot\mat{\mathcal{C}_t & \mathcal{C}_u} = \mat{2 & -\frac{2}{3} \\ -\frac{2}{3} & 2}.
\end{equation}

We now discuss the issue of numerical instabilities and how to detect them. 
This is most relevant to the phase space integration of the virtual contribution, which 
shows numerical cancellations in the calculation of one-loop tensor integrals. Our solution 
is described as follows. 
We use the Ward identities obtained by replacing an effective
current with the corresponding momentum to relate $N$-point integrals to lower point integrals. 
Those identities are called gauge tests and are checked for every phase space point with a small additional computing cost 
by using a cache system. 
The specific tests for the bosonic contributions can be found
in~\bib{Campanario:2011cs}. Here, we concentrate on the hexbox contributions which are most 
complicated. 
The hexbox contributions vanish under the replacement
$J_i(q_i) \to q_i$, because they form EW gauge invariant subsets. 
The gauge structure of these contributions is very rich. 
The subset of three diagrams of \fig{fig:hexbox2} (ignoring the dots), where the position of
the external gauge boson in the upper quark line is fixed and all possible insertions in the lower line
are considered, vanishes for $J_2(q_2) \to
q_2$. Totally, we can construct up to 6 of such identities (three for the
upper and three for the lower line), which are used to flag possible instabilities.

If a bad phase-space point is identified, {\it i.e.} the gauge tests are true
by less than $2$ digits with double precision, the point is discarded. 
For a typical calculation with the inclusive cuts specified below, the 
number of discarded points is statistically negligible. 
This strategy was also successfully applied for $W\gamma\gamma + \text{jet}$
and EW $Hjjj$ production at
NLO QCD in Refs.~\cite{Campanario:2011ud,Campanario:2013fsa} without further need of using any additional rescue system.

Finally, we have a few comments on code optimization and running time. 
Since the leptonic decays of the EW gauge bosons are common for all subprocesses, 
the {\texttt{VBFNLO}} approach is to calculate these decays once for 
each phase-space point and store them. 
Due to the large number of subprocesses, 
we extend this procedure and also precalculate parts of Feynman diagrams, that are common to the subprocesses of the real emission. 
In addition, a caching system to reuse Born amplitudes for 
different dipole terms~\cite{Catani:1996vz} has been implemented. 
With this method, we obtain the NLO inclusive cross section with statistical error of $1\%$ in half an hour 
on an Intel $i5$-$3470$ computer with one core and using the compiler Intel-ifort version $12.1.0$. 

\section{Numerical results}
\label{sec:results}
We choose $M_W=80.385 \GeV$, $M_Z=91.1876 \GeV$, $M_H = 126 \GeV$ and $G_F=1.16637\times 10^{-5}\GeV^{-2}$ as EW input parameters and use the MSTW2008 parton distribution functions~\cite{Martin:2009iq} with $\alpha_s^\text{LO}(M_Z)=0.13939$ and $\alpha_s^\text{NLO}(M_Z)=0.12018$.
Quark mixing effects are neglected and all the fermions, except the top quark with $m_t=173.1 \GeV$, are treated as massless. From this, we get $\Gamma_{W}=2.09761 \GeV$, $\Gamma_Z=2.5089 \GeV$ and $\Gamma_H = 4.195 \MeV$. 
We work in the five-flavor scheme and use the 
$\overline{MS}$ renormalization of the strong coupling constant with the top quark 
decoupled from the running of $\alpha_s$. 
However, the top-loop contribution is explicitly included in the virtual amplitude. 
Subprocesses with external third generation quarks 
should be treated as different processes and are therefore excluded.

We define the inclusive cuts as follows 
\begin{equation}
  \begin{split}
  p_{T(j,l)} > 20 \GeV \qquad \slashed p_T > 30 \GeV \\
  |y_j| < 4.5 \qquad |y_l| < 2.5\qquad R_{l(l,j)} > 0.4 .
\end{split}
\end{equation}
The jets are clustered using the anti-$k_t$ algorithm~\cite{Cacciari:2008gp} with a cone radius of $R=0.4$. 
We use a dynamical factorization and renormalization scale with the central value
\bea
\mu_{0}=\frac{1}{2}
\left(\sum_{\text{partons}} p_{T,i} + 
\sum_{W_i}\sqrt{p_{T,i}^2+m_{W,i}^2}\right),
\label{eq:define_mu0}
\eea
where $m_{W,i}$ denotes the invariant mass of the corresponding leptons. 
This is our default scale choice if not otherwise stated. 

\subsection{Full LO results}
\begin{figure}[t]
  \centering
  \includegraphics[width=0.8\columnwidth]{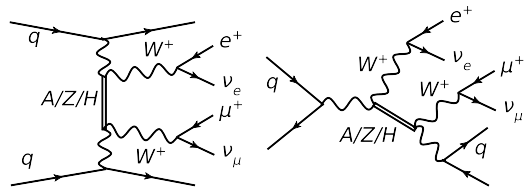}
  \caption{Representative tree-level Feynman diagrams of the EW-induced mechanisms 
for the process $pp\rightarrow e^+\nu_e \mu^+ \nu_\mu~jj$. The double lines represent either a neutral 
EW gauge boson or the Higgs.}
\label{fig:Born_EW_full}
\end{figure}
\begin{figure*}[t]
  \centering
  \includegraphics[width=0.83\columnwidth]{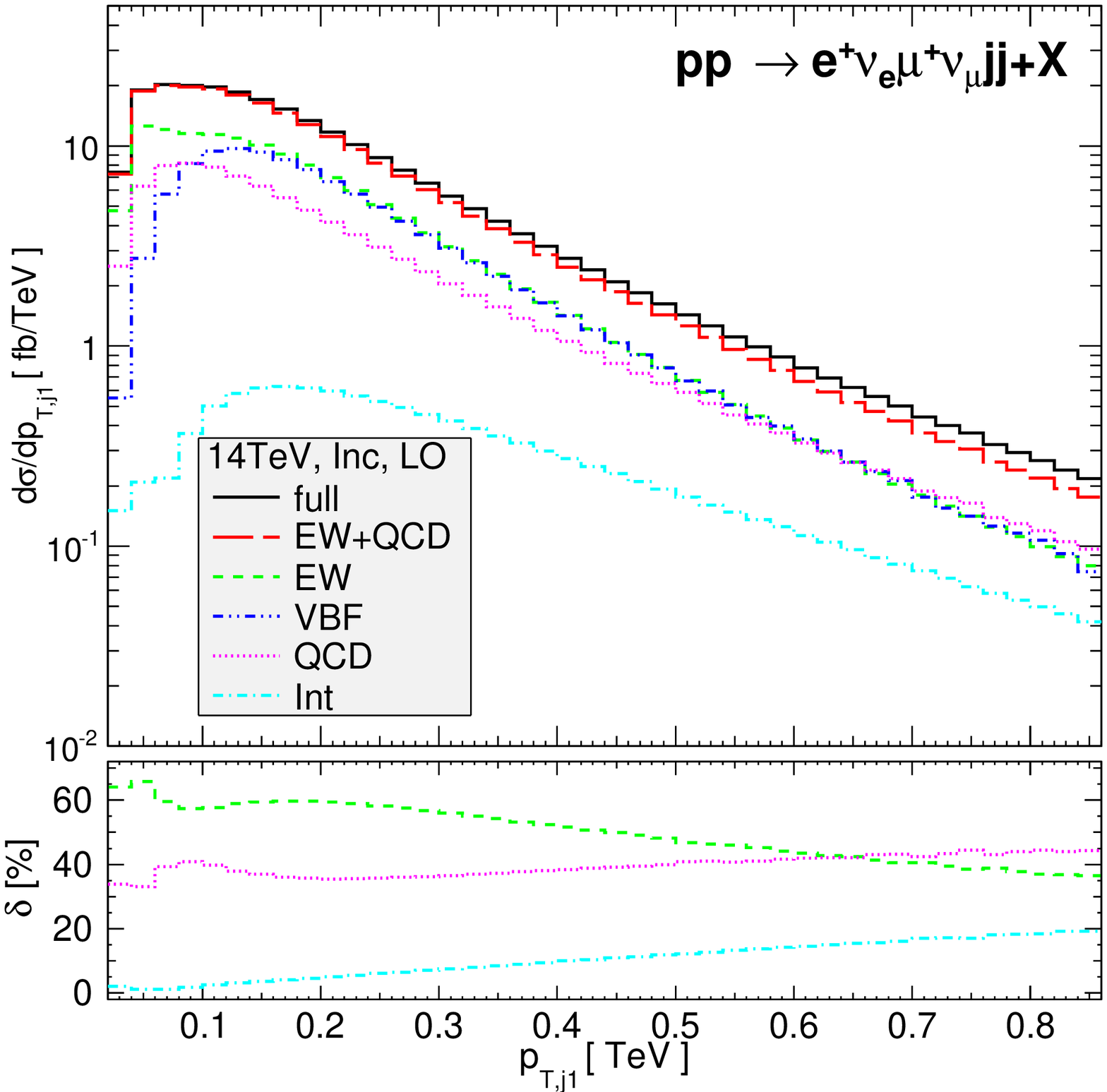}
  \includegraphics[width=0.83\columnwidth]{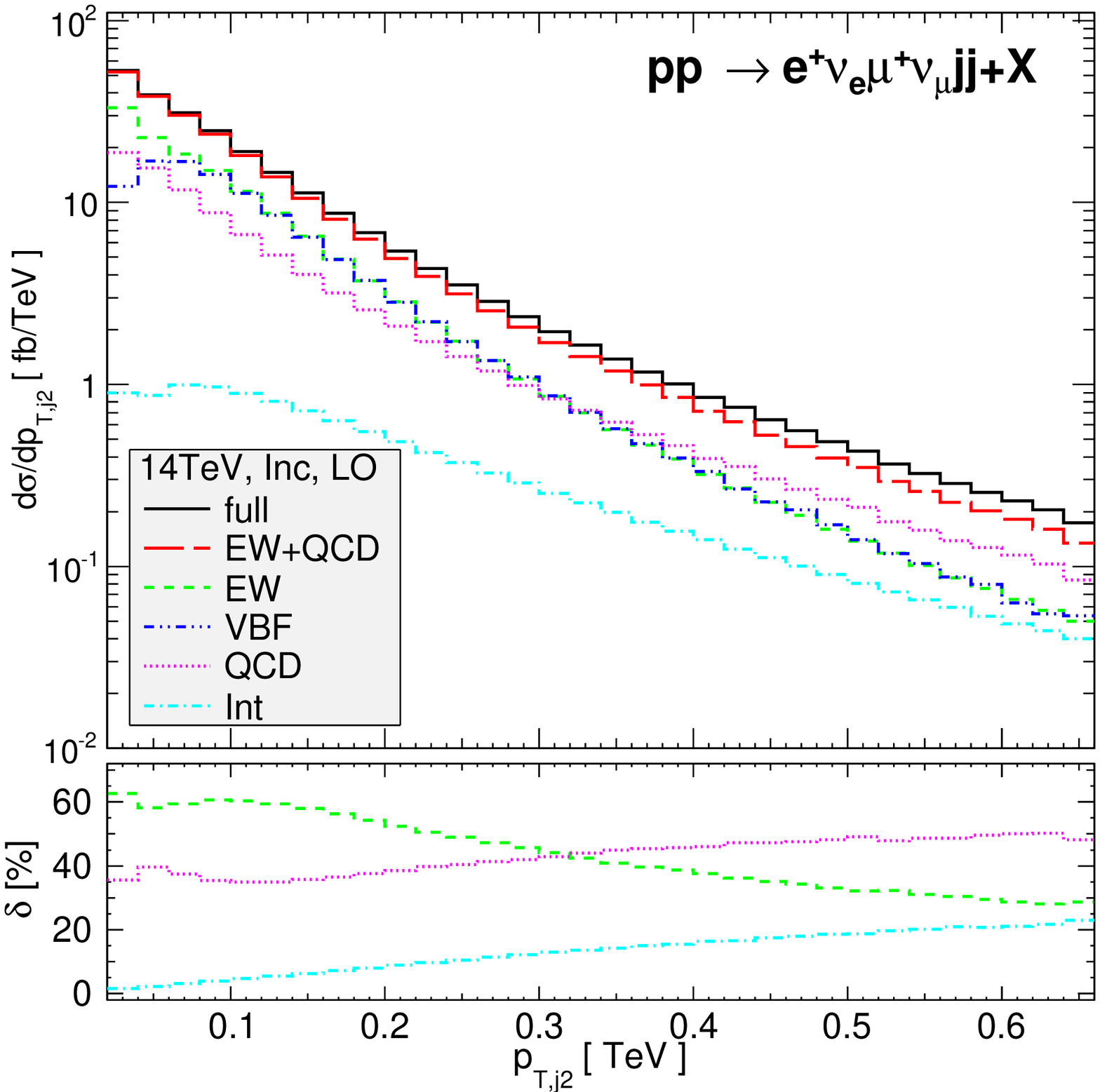}
  \includegraphics[width=0.83\columnwidth]{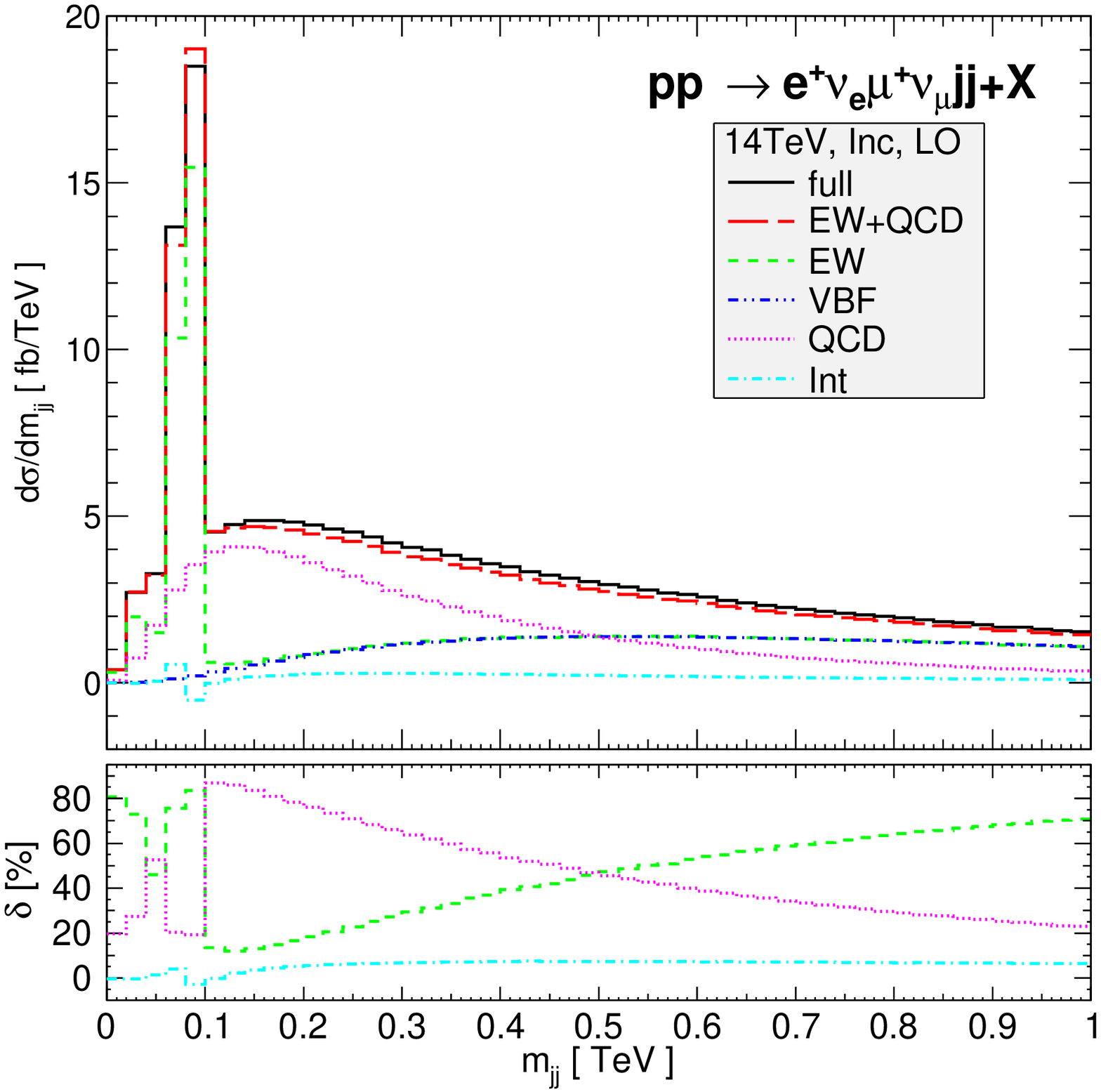}
  \includegraphics[width=0.83\columnwidth]{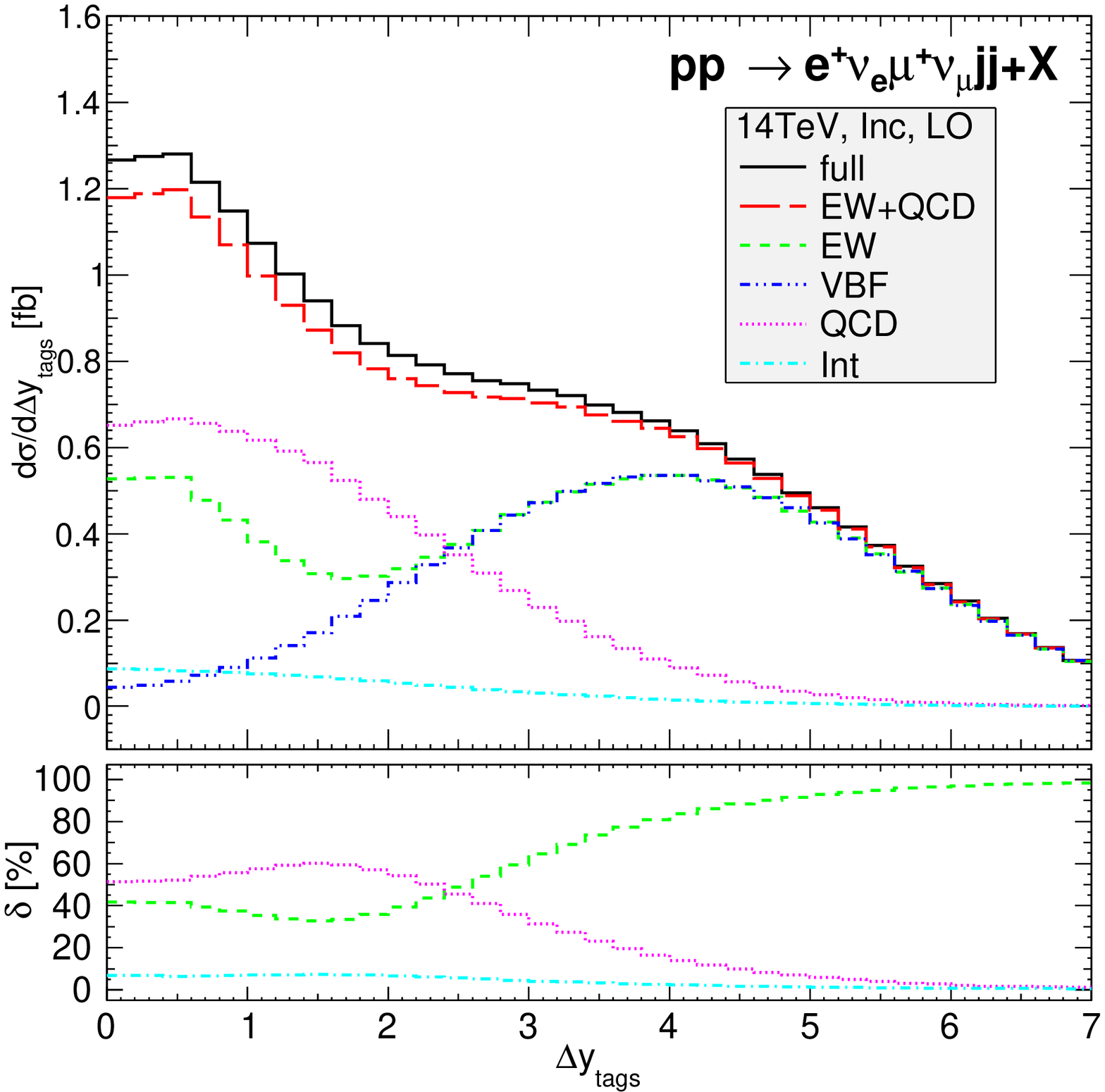}
  \caption{Differential cross sections with inclusive cuts for the transverse momenta (top row) and 
the invariant mass (bottom left) of the two tagging jets ordered by $p_T$. 
The distributions of the rapidity separation between the two jets are in the bottom right panel. 
The relative EW, QCD and interference contributions compared to the full LO results are also plotted in the small panels.
}
\label{dist_LO_jets_inc}
\end{figure*}
We consider the full LO contribution including both the EW-induced and the QCD-induced channels 
and ask the following questions: (i) are the interference effects between the 
EW-induced and the QCD-induced mechanisms important? (ii) what are the cuts for the VBF contribution 
to be dominant? To answer these questions, the full LO cross section is divided into various contributions: 
the QCD-induced contribution of order $\order{\alpha_s^2 \alpha^4}$, the full EW-induced contribution 
including the $t$, $u$ and $s$ channel diagrams as shown in \fig{fig:Born_EW_full} 
of order $\order{\alpha^6}$ and the interference contribution of order $\order{\alpha_s \alpha^5}$. 
The VBF contribution of order $\order{\alpha^6}$ includes only $t$ and $u$ channels. 
The VBF approximation, as implemented in the {\texttt{VBFNLO}} program, neglects 
the interference effects between the $t$ and $u$ channels for processes with two identical quark lines. 
This has been shown to be a very good approximation if VBF cuts are applied, see e.g. \bib{Denner:2012dz}. 
Hereafter, results of the VBF contribution are calculated using the VBF approximation. 
In \fig{dist_LO_jets_inc} we show 
those various contributions as functions of the transverse momentum (top row panels) 
and invariant mass (bottom left panel) of the two tagging jets ordered by $p_T$. 
The distributions of the rapidity separation between the two jets are also displayed in the bottom right panel. 
Here we define $\Delta y_\text{tags} = |y_{j1} - y_{j2}|$. 
The relative EW, QCD and interference contributions compared to the full LO results are also plotted in the small panels.
Our results obtained by using the {\texttt{VBFNLO}} framework have been cross checked against the program 
{\texttt{Sherpa}}~\cite{Gleisberg:2008ta} using the {\texttt{Comix}}~\cite{Gleisberg:2008fv} matrix element generator.
Results at the cross section level for those contributions with different cuts are given in 
\tab{table_onlyLO}. The loose VBF cuts include the inclusive cuts and two additional cuts 
\bea
m_{jj} > 200\GeV,\;\;\; \Delta y_\text{tags} > 2.5.
\label{loose_VBF_cuts}
\eea
The tight VBF cuts are the inclusive cuts together with 
\bea
m_{jj} > 500\GeV,\;\;\; \Delta y_\text{tags} > 4,\;\;\; y_{j1}\cdot y_{j2}<0, 
\label{tight_VBF_cuts}
\eea
and the charged leptons must be in the rapidity gap of the two tagging jets. 
\begin{table}[h]
 \begin{footnotesize}
 \begin{center}
 \caption{\label{table_onlyLO}{LO cross sections (in fb) for various contributions with different cuts as defined in the text. The statistical errors are below 1\,ab.}}
\begin{tabular}{|l|c | r@{.}l | r@{.}l | r@{.}l | r@{.}l | r@{.}l |}
\hline
\multicolumn{2}{|c|}{} &\multicolumn{2}{c|}{VBF}
&\multicolumn{2}{c|}{EW}
&\multicolumn{2}{c|}{QCD}
&\multicolumn{2}{c|}{Int}
&\multicolumn{2}{c|}{Full}
\\
\hline
\multicolumn{2}{|l|}{Inclusive}&  2&189 & 2&784 & 1&810 & 0&234 & 4&828 \\
\multicolumn{2}{|l|}{Loose VBF}&  1&784  & 1&783 & 0&362 & 0&058 & 2&203 \\
\multicolumn{2}{|l|}{Tight VBF}&  0&971  & 0&970 & 0&040 & 0&013 & 1&023 \\
\hline
\end{tabular}\end{center}
 \end{footnotesize}
\end{table}

The numerical results for the inclusive cuts tell us that the interference effect between the QCD and EW induced channels 
is largest when the jets are produced with large $p_T$ and when $\Delta y_\text{tags}$ is small. 
The transverse momentum distributions show a steady increase of this effect, reaching about $20\%$ for
$p_{T,j1}$ about $800\GeV$ or $p_{T,j2}$ about $600\GeV$. 
It reduces to below $10\%$ ($3\%$) for the loose (tight) VBF cuts. 
For the $\Delta y_\text{tags}$ distribution, 
the effect is almost constant for small separation, about $6\%$, and then gradually decreases for 
$\Delta y_\text{tags} > 2$. The interference effect is more democratic in the $m_{jj}$ distribution, 
about $6\%$ for a large range of $m_{jj} > 300\GeV$. We have also looked at the $p_{T,l}$, with $l=e^+,\mu^+$, 
and the $m_{ll}$ distributions (not shown) and observed that the effect is always smaller than $6\%$, being 
largest in the low energy regime. Thus, the interference effect can be large in the $p_{T,j}$ distributions 
at high transverse momentum. However, in this phase space region, the cross section is suppressed. 
Moreover, there is another well-known effect of EW Sudakov corrections due to the exchange of 
a massive gauge boson in loop diagrams, which can introduce negative corrections of about 
$-5\%$ at $p_{T,j}\approx 800\GeV$, see e.g. \bib{Mishra:2013una}.  

\fig{dist_LO_jets_inc} and \tab{table_onlyLO} show also other interesting features of the QCD and EW mechanisms. 
Their contributions are of the same level despite the hierarchy of the coupling constants. The reason is that 
the EW mechanisms are more dynamically enhanced compared to the QCD mechanism. 
For instance, the $s$ channel 
diagram in \fig{fig:Born_EW_full} can have simultaneously three resonating $W$ propagators, dominating 
in the region of small $m_{jj}\approx 80\GeV$. 
The VBF contribution including only $t$ and $u$ channels dominates at 
large $m_{jj}$ and large $\Delta y_\text{tags}$. 
For inclusive cross section, the VBF contribution is slightly larger than 
the QCD one. This is because initial and final state $W$ emissions interfere destructively 
for QCD-induced processes in the central region, 
which is kinematically favored due to the substantial mass of the $W$ bosons. 
This suppression does not occur in the VBF channels because of an additional sign flip in 
the EW charge between the initial and final state $W$ emissions. 
Therefore, if we want to observe the VBF signature, then we can 
impose loose or tight VBF cuts as defined in \eq{loose_VBF_cuts} and \eq{tight_VBF_cuts}, respectively. 
The QCD contribution to the cross section reduces from $37\%$ for the inclusive cuts to 
$16\%$ ($4\%$) for loose (tight) VBF cuts. 
The effects of the VBF cuts are twofold: reducing the QCD-induced contribution and improving the 
VBF approximation. 
With these cuts, we can obtain a very good prediction using the VBF approximation at NLO QCD, the 
QCD-induced contribution at NLO QCD and the interference terms calculated at LO. 
However, if one wants to have more events with same-charge $W$ 
bosons, then measurements with inclusive cuts should also be considered. In this context, NLO QCD corrections 
to the QCD-induced channels are more important and this is the topic of the next section. 
We note that the full EW contribution at LO has been calculated in \bib{Denner:2012dz} with slightly different VBF 
cuts and a different scale choice. Despite these differences, we can see that their results are consistent with ours. 
\begin{figure}[ht]
  \centering
  \includegraphics[width=\columnwidth]{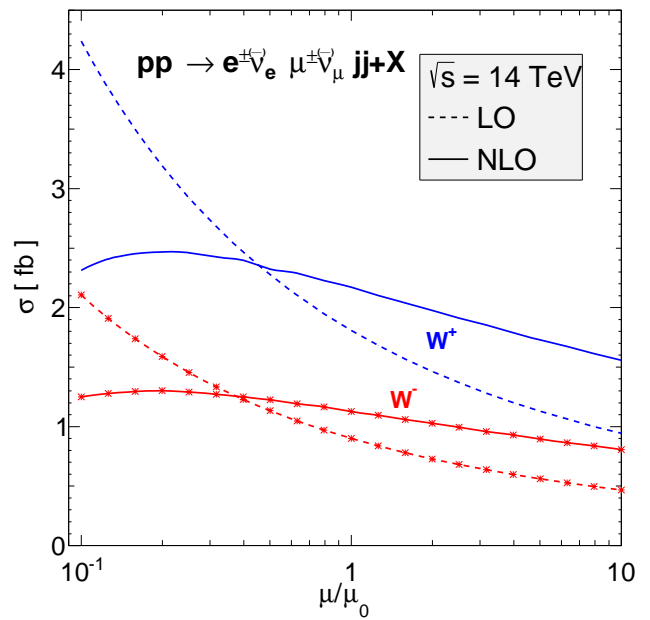}
  \caption{Scale dependence of the LO and NLO cross sections at the LHC. 
The curves with and without stars are for $W^-W^-jj$ and $W^+W^+jj$ productions, respectively.}
  \label{fig:scale}
\end{figure}
\begin{figure*}[ht!]
  \centering
  \includegraphics[width=0.83\columnwidth]{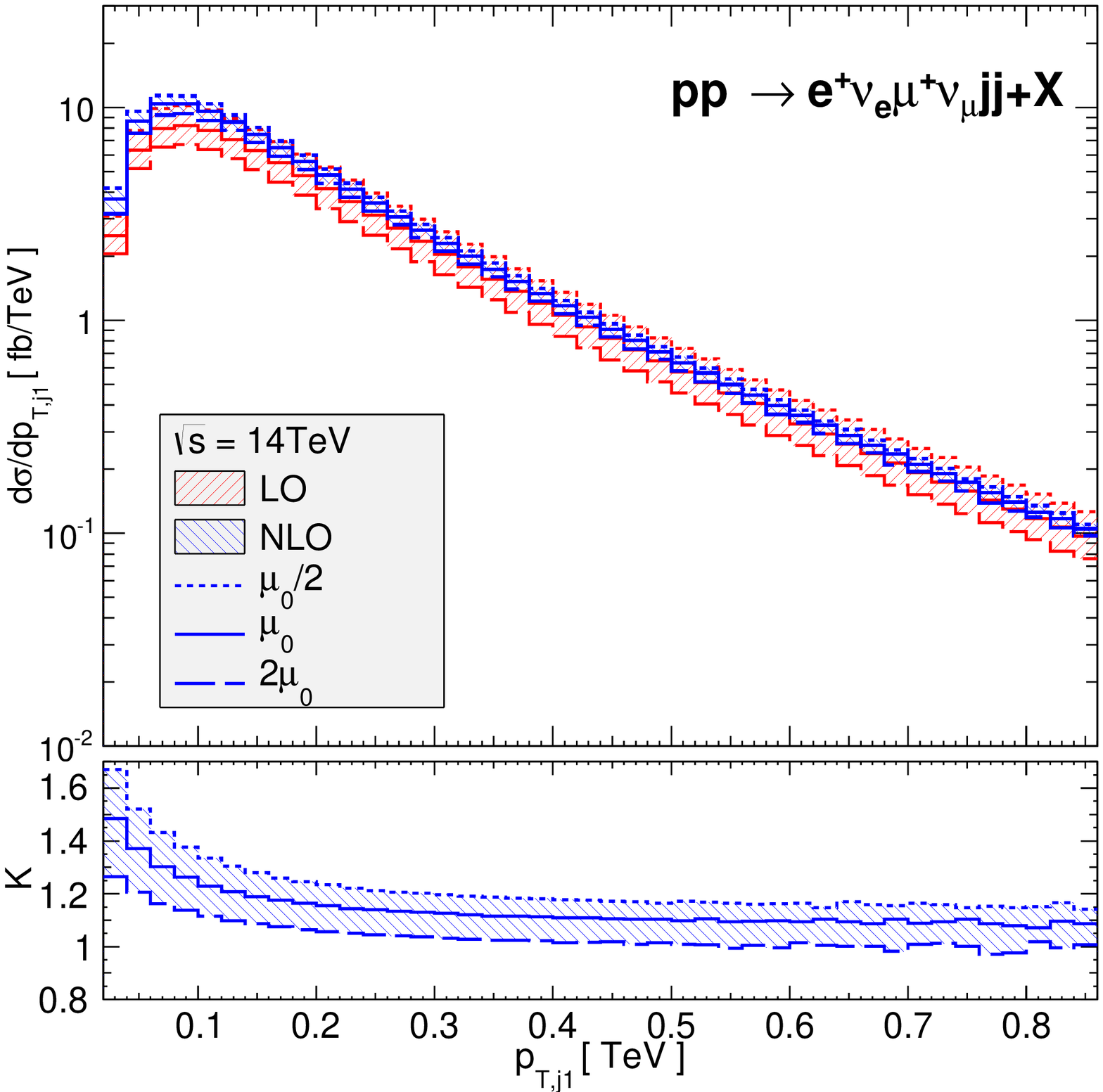}
  \includegraphics[width=0.83\columnwidth]{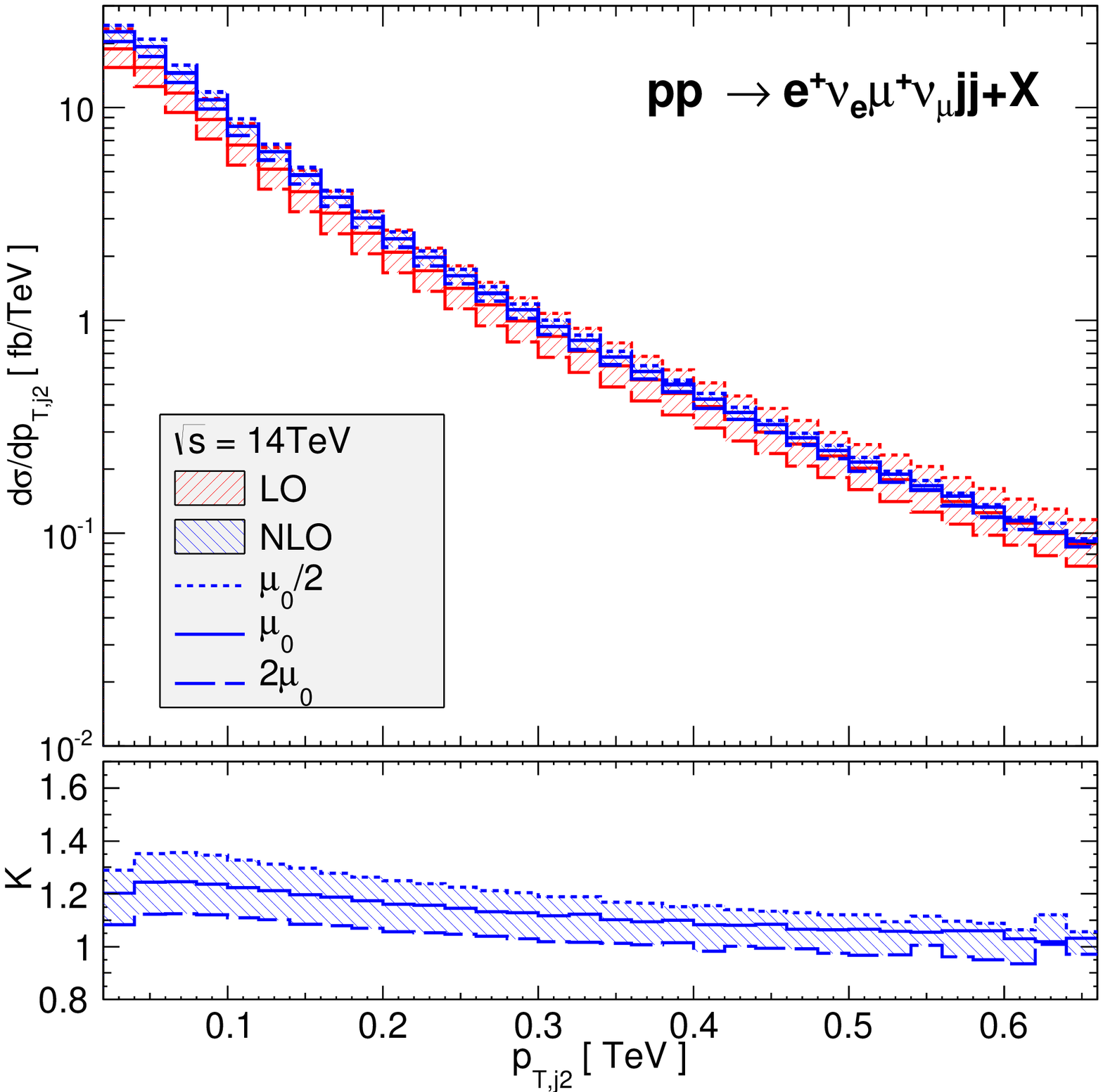}
  \includegraphics[width=0.83\columnwidth]{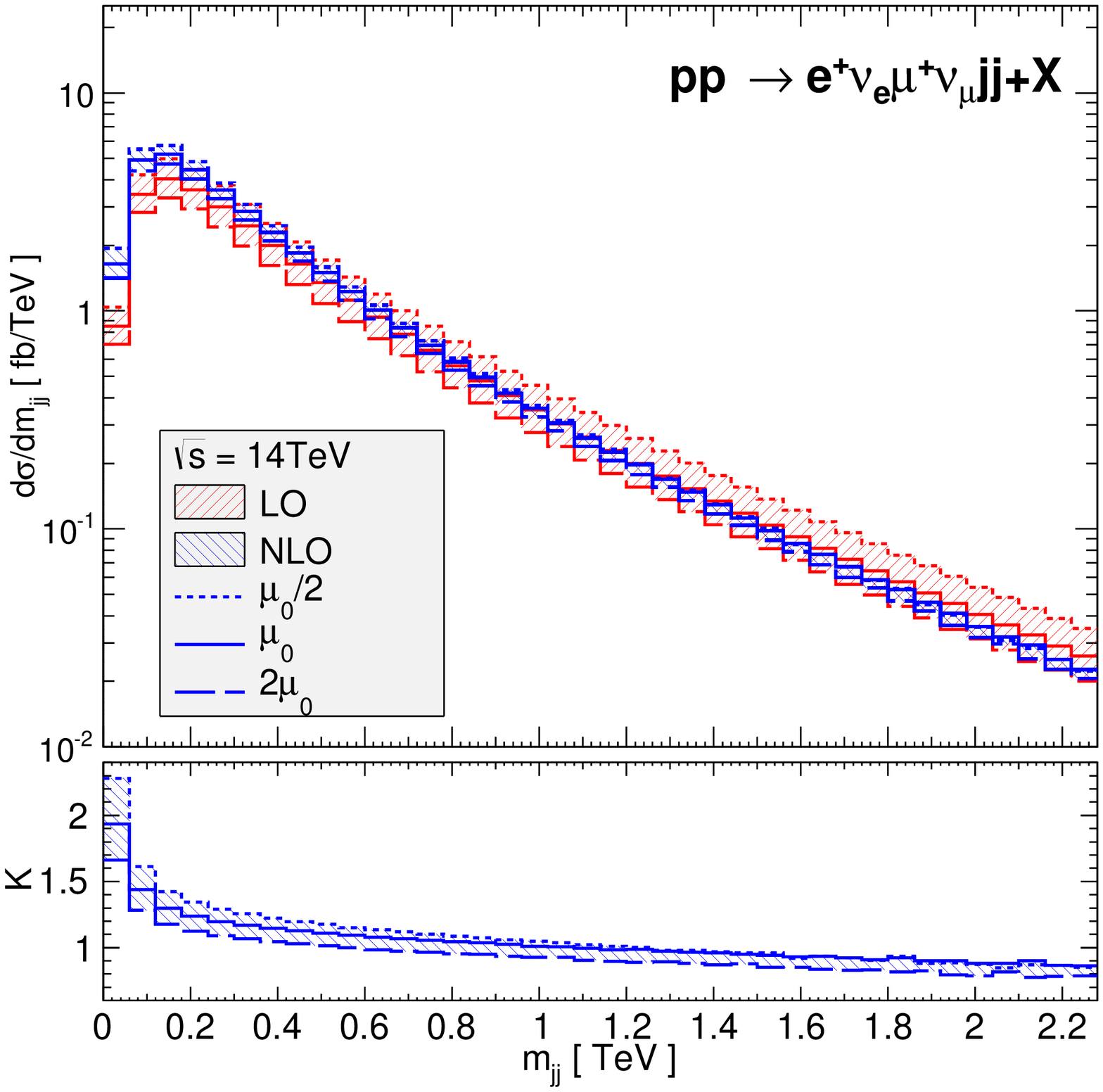}
  \includegraphics[width=0.83\columnwidth]{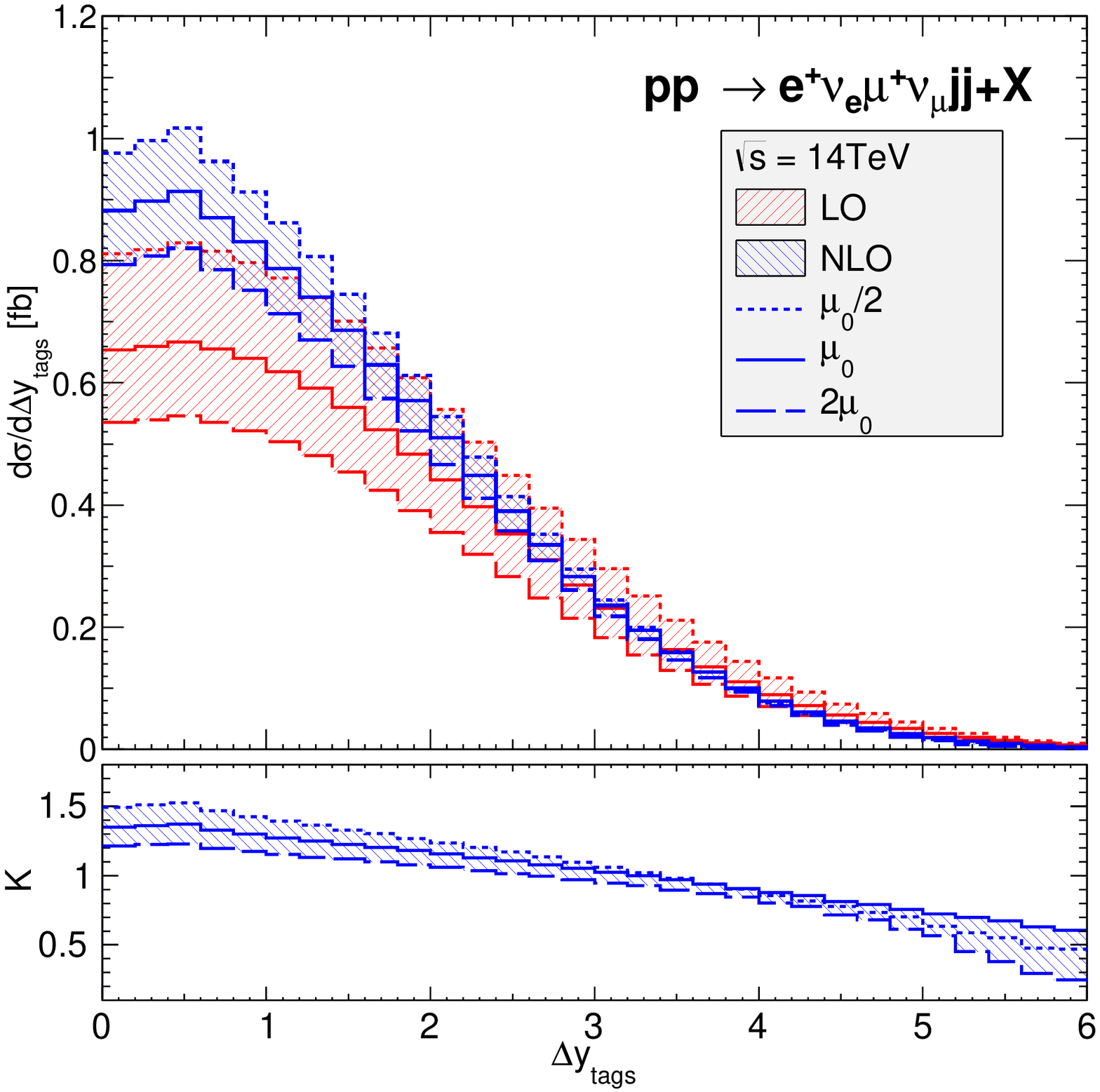}
  \caption{Differential cross sections, for the QCD-induced channels at LO and NLO, 
with inclusive cuts for the transverse momenta (top row) and 
the invariant mass (bottom left) of the two tagging jets ordered by $p_T$. 
The distributions of the rapidity separation between the two jets are in the bottom right panel. 
The bands describe $\mu_0/2 \le \mu_F=\mu_R\le 2\mu_0$ variations. 
The $K$-factor bands are due to the scale variations of the NLO results, 
with respect to $\sigma_\text{LO}(\mu_0)$. 
The solid lines are for the central scale while the dotted and dashed lines correspond to $\mu_0/2$ and $2\mu_0$, respectively.
}
\label{dist_NLO_jets_inc}
\end{figure*}

\subsection{NLO QCD results for the QCD-induced channels}
We now focus on the inclusive cuts and study the NLO QCD corrections to the QCD-induced channels. 
The dependence of the cross section on the scales $\mu_F$ and $\mu_R$, 
which are set equal for simplicity, is shown in \fig{fig:scale}. 
The central value $\mu_0$ is defined in \eq{eq:define_mu0}. 
As expected, we observe a significant reduction in the scale dependence around 
$\mu_0$ when the NLO contribution is included. The uncertainties 
obtained by varying $\mu_{F,R}$ by factors $1/2$ and $2$ around the central value 
are $45\%$ ($45\%$) at LO and $16\%$ ($18\%$) at NLO for the $W^+W^+$ ($W^-W^-$) channel.  
At $\mu = \mu_0$, we get $\sigma_\text{LO}=1.81^{+0.47}_{-0.34}\fb$ ($0.90^{+0.23}_{-0.17}\fb$) 
and $\sigma_\text{NLO}=2.17^{+0.15}_{-0.19}\fb$ ($1.13^{+0.10}_{-0.10}\fb$). 
By varying the two scales separately, we observe a small dependence on $\mu_F$, while 
the $\mu_R$ dependence is similar to the behavior shown in \fig{fig:scale}. 

To understand the phase space dependence of the NLO QCD corrections, we plot, for 
the $W^+W^+$ case, in \fig{dist_NLO_jets_inc} the differential cross sections 
for the transverse momenta (top row) and 
the invariant mass (bottom left) of the two tagging jets ordered by $p_T$ as in the previous section. 
The distributions of the rapidity separation between the two jets are in the bottom right panel. 
The $K$ factors, defined as the ratio of the NLO to the LO results, are 
shown in the small panels. 
The distributions at NLO are much less sensitive
to the variation of the scales than at LO. 
We observe non-trivial phase space dependence of the $K$ factors. 
It varies, for $\mu = \mu_0$, from $1.5$ to $1.1$ for the $p_{T}$ distribution of the hardest jet 
and from $1.2$ to $1$ for the second hardest jet in a large energy range up to about $700 \GeV$. 

A more striking dependence is found in the $\Delta y_\text{tags}$ distributions. 
The $K$ factor is about $1.5$ at $\Delta y_\text{tags} = 0.5$ and decreases rapidly with large 
rapidity separation between the two tagging jets. The QCD correction changes sign at 
about $\Delta y_\text{tags} = 3.3$ and the scale uncertainty band of the $K$ factor has a minimum width at 
about $4$, blowing up with further increasing jet separation. 
This is probably because the invariant mass $m_{jj}$ rapidly increases at large $\Delta y_\text{tags}$, while the tagging jets are mainly produced with low $p_T$, 
\bea
m^2_{jj} \approx 2p_{T,j1}p_{T,j2}[\cosh(y_{j1}-y_{j2}) - \cos(\phi_{j1}-\phi_{j2})].
\label{invMass}
\eea
Note that $\cosh(\Delta y)$ inflates from 1 at $\Delta y=0$ to 27 at $\Delta y=4$.
The low value of $p_{T,j2}$ basically introduces a jet veto for further jet activities. This large difference between the values of $m_{jj}$ and $p_{T,\text{veto}}$
leads to large QCD corrections. 
It then becomes clear that the dynamic scale $\mu_0$ is too small at large $\Delta y_\text{tags}$ and the invariant mass $m_{jj}$ 
should be taken into account in the scale choice. Using a larger scale would make the NLO result more stable and 
reduce the LO result, hence bringing the $K$ factor closer to one as will be shown later.

The $m_{jj}$ distribution shown in the bottom left plot in \fig{dist_NLO_jets_inc} does not exhibit the same behavior at large $m_{jj}$. The $K$ factor is about 
one at $1\TeV$ and the uncertainty band is regular.
At large $m_{jj}$, the dominant contribution comes from the configuration with small 
$\Delta y_\text{tags}$ (see bottom right plot in \fig{dist_NLO_jets_inc}) and large transverse momenta. Therefore, using $\mu_0$ is reasonable here. 
However, an additional feature appears in the low energy regime where a large $K$ factor occurs, 
being larger than $2$ for $m_{jj} < 30 \GeV$. 
At LO, there are two quark jets which are well separated because there is 
a finite IR cutoff due to the $W$ mass. 
The $m_{jj}$ distribution has a peak at about $150\GeV$ at LO. The peak's position is shifted 
to a smaller value at NLO, leading to a large $K$ factor at small $m_{jj}$. 
This is due to a new kinematic configuration opening up at NLO, 
where a final-state quark splits into a quark-gluon pair, resulting in two jets with 
low invariant mass. 
The second quark can either 
be unobserved or be combined in the quark jet. We have explicitly verified this by computing 
the contribution separately. 

\begin{figure*}[ht!]
  \centering
  \includegraphics[width=0.83\columnwidth]{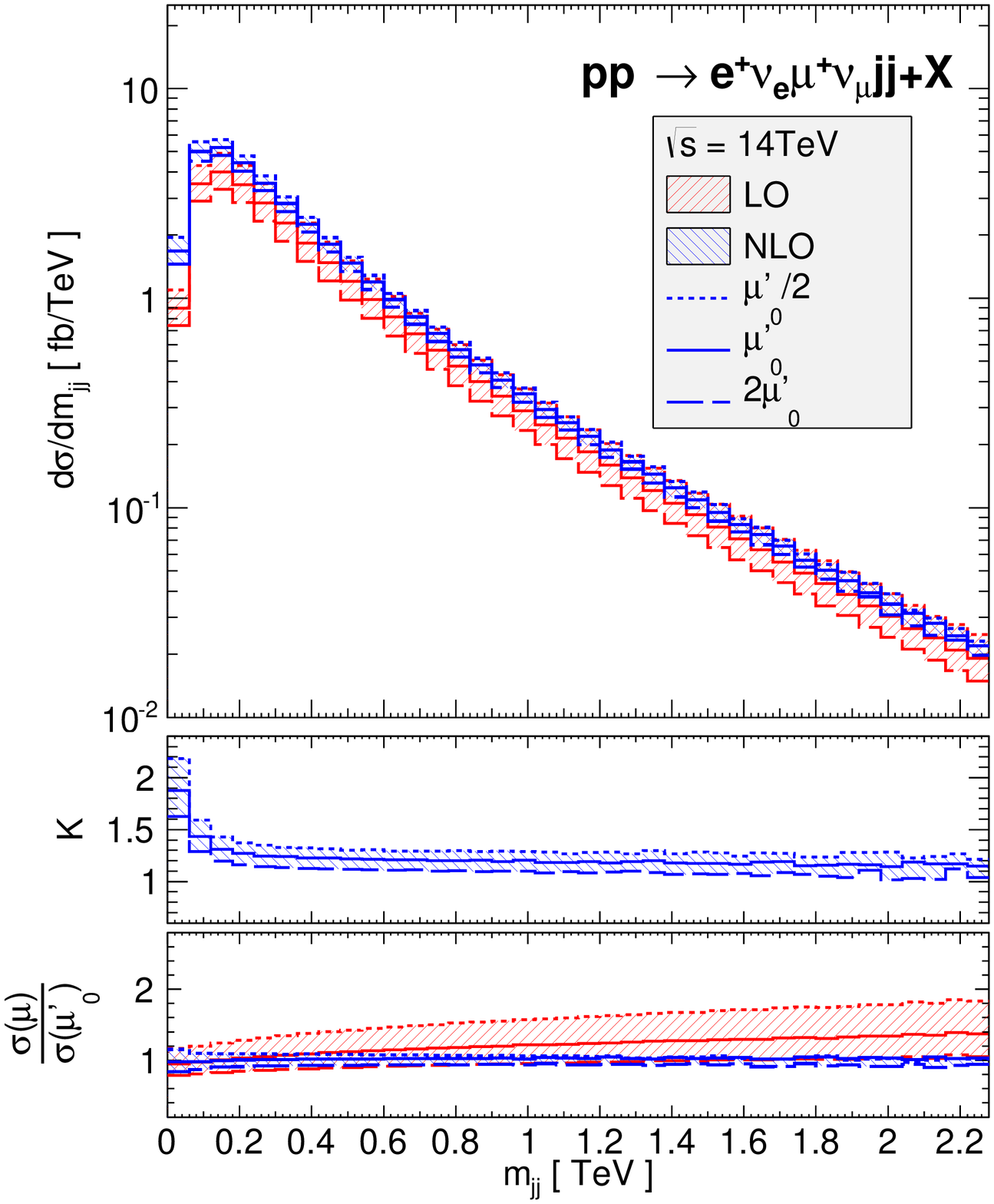}
  \includegraphics[width=0.83\columnwidth]{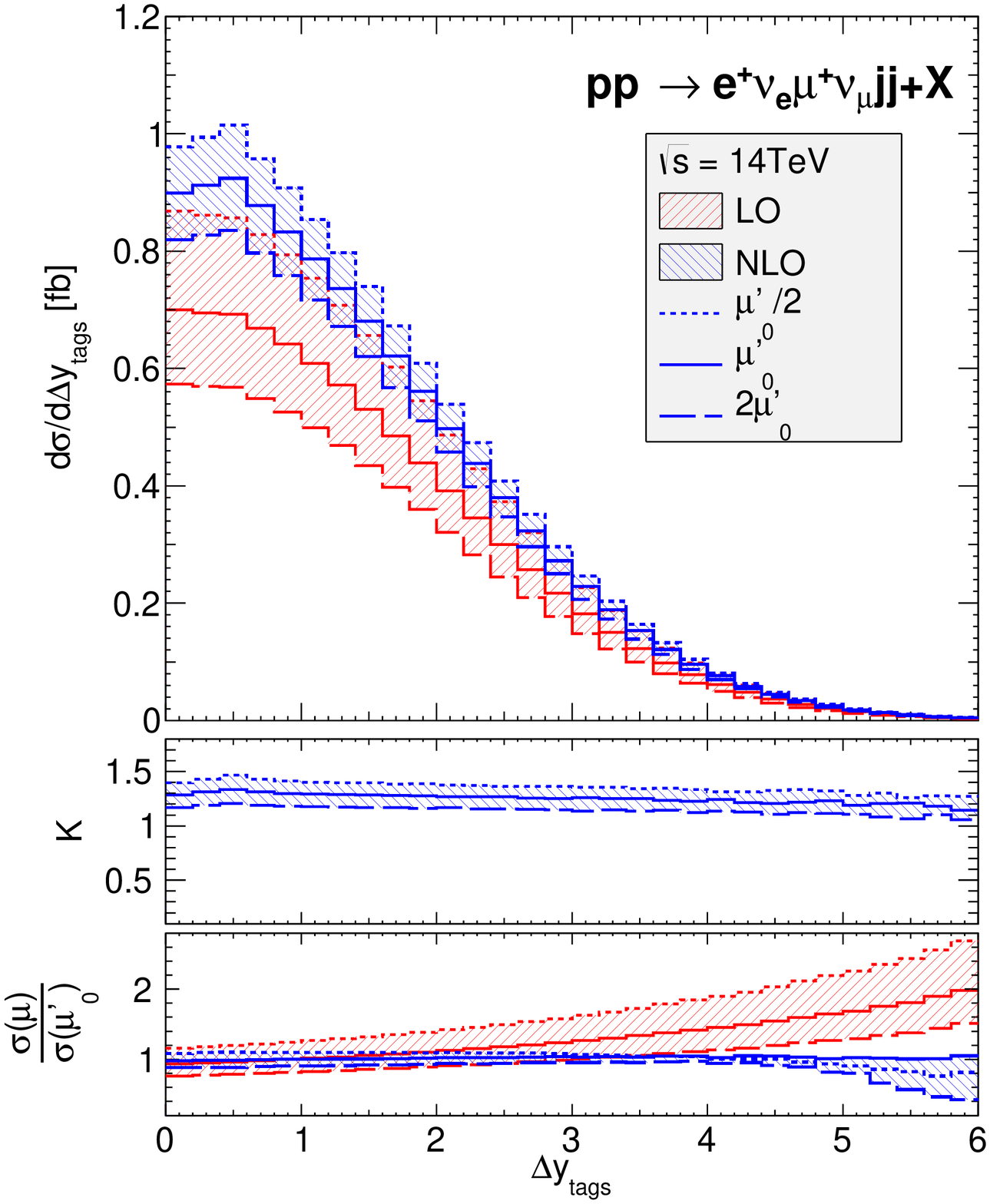}
  \caption{Similar to \fig{dist_NLO_jets_inc} but for the new scale choice 
$\mu_F = \mu_R = \mu^{\prime}_{0}$ defined in \eq{eq:define_mu0_prime}. 
Additional panels at the bottom show the ratios of differential 
cross sections with the default scale $\mu_{0}$ over the ones with $\mu^{\prime}_{0}$ at 
LO and NLO. The bands on these ratios 
show the scale variations $\mu_0/2 < \mu < 2 \mu_0$ of the numerators while the denominators are calculated at 
the central scale. 
}
\label{dist_NLO_jets_inc_prime}
\end{figure*}

The above discussion indicates that the choice of scales is not trivial
and the best choice can be observable dependent, being more important
for LO predictions as we will show. 
A better scale taking into account that
$m_{jj}$  can be much larger than the default scale $\mu_0$, Eq.~\ref{eq:define_mu0}, when 
$\Delta y_\text{tags}$ is large was proposed in \bib{Ellis:1992en} in
the framework of di-jet production. To accommodate the vector bosons, we
slightly modify it and propose 
\bea
\mu^{\prime}_{0}=\frac{1}{2}
\left(\sum_{\text{jets}} p_{T,i} \exp{|y_{i}-y_{12}|} + 
\sum_{W_i}\sqrt{p_{T,i}^2+m_{W,i}^2}\right),
\label{eq:define_mu0_prime}
\eea
where $y_{12} = (y_1 + y_2)/2$ denotes the average
rapidity of the two hardest jets. The first contribution in this scale interpolates between
$\sum_{\text{jets}} p_{T,i}/2$ and $m_{jj}$, so that $\mu_0^{\prime}$ is approximately
equal to $\mu_0$ and $m_{jj}$ for small and large $\Delta y_\text{tags}$ values, respectively. 
New distributions with this scale choice are shown in 
\fig{dist_NLO_jets_inc_prime} for $\Delta y_\text{tags}$ and
$m_{jj}$. The other $p_{T,j}$ distributions 
shown in Fig.~\ref{dist_NLO_jets_inc} are not much 
affected and therefore not shown for the new scale. 
As expected, we observe that the $K$ factor for the $\Delta y_\text{tags}$ distribution 
becomes more regular, flatter and closer to one. 
The $K$ factor for the $m_{jj}$ distribution gets also flatter at large
$m_{jj}$. Furthermore, in the lower panel where the ratio of the
predictions of the two scales at LO and NLO are plotted, one can observe
that the differences are
mainly due to the changes of the LO predictions and the NLO are more
stable. This points out the sensitivity of the LO prediction to different
scale choices and the relevance of the NLO predictions to stabilize the results.
These results justify the above argument about the shortcoming of the scale $\mu_0$. 


\section{Conclusions}
In this paper, we have presented a study of the production of two equal-charge $W$ 
bosons in association with two jets at the LHC with $14 \TeV$ center-of-mass energy. 
The full LO calculation has been done and 
it was found that the interference effects between the QCD-induced and EW-induced channels 
can be large if the jets are produced with high transverse momentum, reaching about $20\%$ 
at $p_{T,j1} \approx 800 \GeV$, for inclusive cuts. It reduces to below $3\%$ for 
tight VBF cuts. The full LO results also show that the QCD-induced contribution 
is largely removed using VBF cuts. 

We have also studied the QCD-induced channels at NLO QCD using the inclusive cuts. 
As expected, it was found that the inclusion of the QCD corrections reduces 
significantly the scale uncertainties of the cross section and of
differential distributions. We have discussed the dependence of the 
QCD corrections on the kinematics of the two tagging jets for two different scale choices.
For the $p_{T,j}$ distributions, the $K$ factor for the central scale varies from $1$ to $1.5$ for a 
large energy range up to $800 \GeV$. 
The $K$ factor of the invariant mass distribution 
is larger than $2$ for $m_{jj} < 30 \GeV$ due to a new kinematic configuration opening up at NLO. 
For large invariant mass, $m_{jj} > 200 \GeV$, the QCD correction is smaller than $20\%$.  
More interesting is the dependence on the rapidity separation between the two hardest jets. 
For our default scale $\mu_0$, given by the sum of jet and $W$ transverse energies, 
the $K$ factor decreases steadily with increasing separation and becomes smaller than one for 
$\Delta y_\text{tags} > 3.3$. The scale uncertainty band has a minimum width at $\Delta y_\text{tags} = 4$ 
and then blows up with increasing jet separation. This is because the central scale $\mu_0$ is too 
small compared to the hard scale $m_{jj}$, which typically is large for high $\Delta y$. This conclusion is 
justified by the results obtained using another scale $\mu^{\prime}_0$, 
which takes into account $m_{jj}$ for large jet separations. 
By choosing $\mu^{\prime}_0$ the LO results are much closer to the NLO
ones and the phase space dependence of the K factors is strongly reduced for the observables shown in this article. 

On a technical note, compared to the first calculation reported in \bib{Melia:2010bm}, 
we allow the intermediate $W$ bosons to be off-shell and the 
top-loop contribution is included. Neglecting these small effects, we obtain a very good agreement with the results 
of \bib{Melia:2010bm}, both at the amplitude squared and cross section levels. 
Finally, our code will be publicly available as part of the {\texttt{VBFNLO}} 
program~\cite{Arnold:2008rz,*Arnold:2012xn}, thereby further studies of the QCD corrections with different kinematic 
cuts can be easily done.

\begin{acknowledgments}
We acknowledge the support from the Deutsche Forschungsgemeinschaft
via the Sonderforschungsbereich/Transregio SFB/TR-9 Computational Particle Physics.
FC is funded by a Marie Curie fellowship (PIEF-GA-2011-298960) and partially by MINECO (FPA2011-23596) and by LHCPhenonet (PITN-GA-2010-264564).  
MK is supported by the Graduiertenkolleg 1694 ``Elementarteilchenphysik bei h\"ochster Energie und h\"ochster Pr\"azision''.
\end{acknowledgments}

\appendix*
\section{Results at one phase-space point}
\label{appendixA}
In this appendix, we provide results at a random phase-space point to facilitate comparisons
with our results. The phase-space point
for the process $q_1 \bar{q}_2 \to \bar{q}_3 q_4 e^+ \nu_e \mu^+ \nu_\mu$ is given in \tab{table_PSP_2to6}.
\begin{table*}[th]
 \begin{footnotesize}
 \begin{center}
 \caption{\label{table_PSP_2to6}{Momenta (in GeV) at a random phase-space point for $q_1 \bar{q}_2 \to \bar{q}_3 q_4 e^+ \nu_e \mu^+ \nu_\mu$ subprocesses.}}
\begin{tabular}{l | r@{.}l r@{.}l r@{.}l r@{.}l}
& \multicolumn{2}{c}{ $E$}
& \multicolumn{2}{c}{ $p_x$}
& \multicolumn{2}{c}{ $p_y$}
& \multicolumn{2}{c}{ $p_z$}
\\
\hline
$q_1$  & 18&3459102072588 & 0&0  & 0&0  &  18&3459102072588  \\
$\bar{q}_2$  & 4853&43796816526 & 0&0  & 0&0  &  -4853&43796816526  \\
$\bar{q}_3$  & 235&795970274883 & -57&9468743482139 & -7&096445419113396$\times 10^{-15}$ & -228&564869022223 \\
$q_4$  & 141&477229270568 & -45&5048903376581 & -65&9221967646567 & -116&616359620580 \\
$e^+$  & 276&004829895761 & 31&4878768361538 & -8&65306166938040 & -274&066240646098 \\
$\nu_e$  & 1909&28515244344 & 29&6334571080402 & 40&1409467910328 & -1908&63311192893 \\
$\mu^+$  & 2241&46026948104 & 28&1723094714198 & 30&2470561132914 & -2241&07910976778 \\
$\nu_\mu$  & 67&7604270068059 & 14&1581212702582 & 4&18725552971283 & -66&1323669723852 \\
\hline
\end{tabular}\end{center}
 \end{footnotesize}
\end{table*}
\begin{table*}[th]
 \begin{footnotesize}
 \begin{center}
\caption{\label{table_PSP_QCD_udud}{QCD interference amplitudes $2\text{Re}(\mathcal{A}_\text{NLO}\mathcal{A}^{*}_\text{LO})$
for $u \bar{d} \to \bar{u} d e^+ \nu_e \mu^+ \nu_\mu$ subprocess.}}
\begin{tabular}{l | r@{.}l r@{.}l r@{.}l}
& \multicolumn{2}{c}{ $1/\epsilon^2$}
& \multicolumn{2}{c}{ $1/\epsilon$}
& \multicolumn{2}{c}{ finite}
\\
\hline
I operator  & 1&053330833015670$\times 10^{-2}$ & -5&936404140457268$\times 10^{-3}$  & 2&640293552481222$\times 10^{-3}$ \\
loop  & -1&053330833914469$\times 10^{-2}$ & 5&936404145582775$\times 10^{-3}$ & 8&079973631744798$\times 10^{-3}$  \\
I+loop  &  -8&987987823141610$\times 10^{-12}$ & 5&125507271694671$\times 10^{-12}$ & 1&072026718422602$\times 10^{-2}$ \\
\hline
\end{tabular}\end{center}
 \end{footnotesize}
\end{table*}
\begin{table*}[th!]
 \begin{footnotesize}
 \begin{center}
 \caption{\label{table_PSP_QCD_udcs}{QCD interference amplitudes $2\text{Re}(\mathcal{A}_\text{NLO}\mathcal{A}^{*}_\text{LO})$
for $u \bar{d} \to \bar{c} s e^+ \nu_e \mu^+ \nu_\mu$ subprocess.}}
\begin{tabular}{l | r@{.}l r@{.}l r@{.}l}
& \multicolumn{2}{c}{ $1/\epsilon^2$}
& \multicolumn{2}{c}{ $1/\epsilon$}
& \multicolumn{2}{c}{ finite}
\\
\hline
I operator  &  3&455808248747951$\times 10^{-3}$ & -1&006571544788643$\times 10^{-4}$  & -1&947046109412057$\times 10^{-3}$ \\
loop  & -3&455808251685664$\times 10^{-3}$ & 1&006571546305143$\times 10^{-4}$ & -5&605259390551837$\times 10^{-3}$ \\
I+loop  & -2&937712573203299$\times 10^{-12}$ & 1&516500563242879$\times 10^{-13}$ & -7&552305499963894$\times 10^{-3}$ \\
\hline
\end{tabular}\end{center}
 \end{footnotesize}
\end{table*}
In the following we provide the squared amplitude averaged over the initial-state 
helicities and colors. We also set $\alpha = \alpha_s = 1$ for simplicity. 
The top quark is decoupled from the running of $\alpha_s$. However, its contribution 
is explicitly included in the one-loop amplitudes. At tree level, we have
\begin{align}
  \overline{|\mathcal{A}_\text{LO}^{u\bar{d}\rightarrow \bar{u}d}|}^2 &= 1.240926153611845\times 10^{-2},\nonumber \\
  \overline{|\mathcal{A}_\text{LO}^{u\bar{d}\rightarrow \bar{c}s}|}^2 &= 4.071278180896138\times 10^{-3}.
\end{align}
The interference amplitudes
$2\text{Re}(\mathcal{A}_\text{NLO}\mathcal{A}^{*}_\text{LO})$,
for the one-loop corrections (including counterterms) and the I-operator contribution as defined in \bib{Catani:1996vz},
are given in \tab{table_PSP_QCD_udud} and \tab{table_PSP_QCD_udcs}. Here we use the following
convention for the one-loop integrals, with $D=4-2\epsilon$,
\bea
T_0 = \frac{\mu_R^{2\epsilon}\Gamma(1-\epsilon)}{i\pi^{2-\epsilon}}\int d^D q \frac{1}{(q^2 - m_1^2 + i0)\cdots}.
\eea
This amounts to dropping a factor ${(4\pi)^\epsilon}/{\Gamma(1-\epsilon)}$
both in the virtual corrections and the I-operator.
Moreover, the conventional dimensional-regularization method~\cite{'tHooft:1972fi}
with $\mu_{R} = 80\GeV$ is used. 
Changing from the conventional dimensional-regularization method to the dimensional reduction scheme 
induces a finite shift. 
This shift can be easily found by observing that 
the sum $|\mathcal{A}_\text{LO}|^2+2\text{Re}(\mathcal{A}_\text{NLO}\mathcal{A}^{*}_\text{LO})$ 
must be unchanged as explained in \bib{Catani:1996pk}. 
Thus, the shift on $2\text{Re}(\mathcal{A}_\text{NLO}\mathcal{A}^{*}_\text{LO})$ is opposite to the shift 
on the Born amplitude squared, which in turn is given by the following change in the strong coupling constant, 
see e.g. \bib{Kunszt:1993sd},
\bea
\alpha_s^{\overline{DR}} = \alpha_s^{\overline{MS}}\left(1+\frac{\alpha_s}{4\pi}\right).
\eea
The shift on the I-operator contribution can easily be calculated using the rule given in \bib{Catani:1996vz}.

\bibliographystyle{../h-physrev}

\end{document}